\documentclass[12pt,preprint]{aastex}

\def\degree{\ifmmode {^\circ}\else {$^\circ$}\fi}
\def\rstar{\ifmmode {\, R_{\star}}\else $R_{\star}$\fi}
\def\msol{\ifmmode {\, M_{\odot}}\else $M_{\odot}$\fi}
\def\rsol{\ifmmode {\, R_{\odot}}\else $R_{\odot}$\fi}
\def\lsol{\ifmmode {\, L_{\odot}}\else $L_{\odot}$\fi}
\def\msolyr{\ifmmode {\,M_{\odot}\,{\rm yr}^{-1}}\else $M_{\odot}\,{\rm yr}^{-
1}$\fi}
\def\mdot{\ifmmode {\,\dot{M}}\else $\dot{M}$\fi}

\def\mdotyr{\ifmmode {\,\dot{M}\,yr^{-1}}\else $\dot{M}\,yr^{-1}$\fi}

\newcommand{\Vinf}{\mbox{$V_{\infty}$}~}
\newcommand{\kms}{km s$^{-1}$}
\newcommand{\Teff}{\ifmmode{T_{\rm eff}}\else{$T_{\rm eff}$}}

\begin{document}

\title{Winds in R Coronae Borealis Stars}
\author{Geoffrey C. Clayton$^1$, T.R. Geballe$^2$, and Luciana Bianchi$^3$}

\altaffiltext{1}{Department of Physics \& Astronomy, Louisiana State 
University
   Baton Rouge, LA 70803 email: gclayton@fenway.phys.lsu.edu}

\altaffiltext{2}{Gemini Observatory, 670 N. A'ohoku Place, Hilo, HI 96720 
email: tgeballe@gemini.edu }

\altaffiltext{3}{Center for Astrophysical Sciences, The Johns Hopkins University, Department of Physics and Astronomy, 239 Bloomberg Center for Physics and Astronomy, 3400 North Charles Street, Baltimore, MD 21218
email: bianchi@marmolada.pha.jhu.edu}

\begin{abstract}
We present new spectroscopic observations of the He I $\lambda$10830
line in R Coronae Borealis (RCB) stars which
 provide the first strong
 evidence that most, if not
all, RCB stars have winds. 
It has long been suggested that when dust forms around an RCB star, radiation 
pressure accelerates the dust away from the star, dragging the gas along with 
it. 
The new spectra show that nine of the ten stars 
observed have P-Cygni or asymmetric blue-shifted profiles 
in the He I $\lambda$10830 line. 
In all cases, the He I line indicates a mass outflow - with a range
of intensity and velocity.
Around the RCB stars, it is likely that this state is populated
by collisional excitation
rather than photoionization/recombination.
The line profiles have been
modeled with an SEI code to derive the 
optical depth and the velocity field of the helium gas.
The results show that the typical RCB wind has a steep acceleration with
a terminal velocity of
\Vinf = 200-350 \kms~and a column density of 
N $\sim$10$^{12}$ cm$^{-2}$ in the He I $\lambda$10830 line.
There is a possible relationship between the lightcurve of an RCB
star and its He I $\lambda$10830
profile. Stars which have gone hundreds of days with no dust-formation 
episodes tend to
have weaker He I features. The unusual RCB star, V854 Cen, does not follow
this trend, showing little or no He I absorption despite high mass-loss 
activity.
The He I $\lambda$10830 line in R CrB itself,
which has been observed at four epochs between 1978 and 2001, seems to show
a P-Cygni or asymmetric blue-shifted 
profile at all times whether it is in decline or at maximum light.
 
\end{abstract}



\section{Introduction}

The RCB stars are a small group of hydrogen-deficient carbon-rich 
supergiants which undergo spectacular declines in brightness of up to
8 magnitudes
at irregular intervals (Clayton 
1996). 
RCB star atmospheres are extremely deficient in hydrogen but very rich in carbon.
Dust is apparently forming within a couple of stellar radii of the stars, which have 
$T_{eff}\sim5000-7000$~K.
RCB stars are very rare. Only about 35 are known in the Galaxy (Clayton 1996).
Their rarity may stem from the fact that they are in an extremely rapid 
phase of the evolution toward white dwarfs. 
Understanding the RCB stars is a key test
for any theory that aims to explain hydrogen deficiency in
post-Asymptotic Giant Branch stars.  
There are two major evolutionary models for the origin of RCB stars: 
the Double Degenerate and the Final Helium Shell Flash (Iben et al. 1996). 
The former involves the merger of two white dwarfs, and in the latter a 
white dwarf/evolved Planetary Nebula (PN) central star
is blown up to supergiant size by a final helium flash.
In the final flash model, there is a close relationship between RCB stars and
PN. This connection has recently become stronger, 
since 
the central stars of three old PN's (Sakurai's Object, V605 Aql and FG Sge) 
have been observed to undergo final-flash outbursts which transformed them from hot evolved central 
stars into cool giants with the spectral properties of RCB stars
(Kerber et al. 1999; Asplund et al. 1999; Clayton \& De Marco 1997; 
Gonzalez et al. 1998).
Two of these stars, FG Sge and 
Sakurai's Object are in an RCB-like phase at present.

During a decline,  a cloud of 
carbon-rich dust forms along the line of sight, eclipsing the photosphere, and revealing a 
rich emission-line spectrum made up primarily of neutral and singly ionized species.
The emission lines suggest at 
least two temperature regimes; a cool 
($\sim$5000 K) inner region likely to be the 
site of neutral and singly-ionized species producing a narrow-line
spectrum, and a much hotter outer region indicated by the presence 
of broad emission lines such as C III $\lambda$1909, C IV 
$\lambda$1550 and He I $\lambda$10830\footnote{The He I $\lambda$10830 
line is a triplet. The vacuum wavelengths of 
the triplet are 1.083206, 1.083322 and 1.083331 \micron. Vacuum wavelengths are plotted in the figures.  However, 
we will continue with tradition and refer to this as the 
He I $\lambda$10830 line.} (Wing et al. 1972; Querci \& Querci 1978; Zirin 1982; Clayton et al. 1992; Lawson et al. 1999).
Other broad lines, such as 
Na I D and Ca II H \& K, imply a cooler region.  
The possible detection of 
C IV $\lambda$1550 implies the presence of a transition region 
with an electron temperature $T_{\rm e} \sim 10^{5}$ K (Jordan \& 
Linsky 1987). But RCB stars do not exhibit a normal chromospheric spectrum so
the presence of high excitation lines in the cool RCB stars has been 
perplexing.  

He I $\lambda$10830 
was detected in R CrB thirty years ago,  
when it was just below 
maximum light (Wing et al. 1972; Querci \& Querci 1978; Zirin 1982). The line showed a P-Cygni 
profile with a violet displacement of more than 200 km $s^{-1}$. 
This line is similar to that measured for Sakurai's Object in 
its RCB-phase (Eyres et al. 1999). 
Since 1978, the He I $\lambda$10830 line in R CrB has been observed only once.
It was seen strongly in emission while R CrB was 
recovering from a deep decline in 1996 (Rao et al. 1999).
No further observations of He I $\lambda$10830 
in RCB stars exist in the literature.  
 
\section{Observations}
Observations of He I $\lambda$10830 in ten RCB stars 
were obtained on 15 June 2001 at UKIRT using the grating 
spectrometer CGS4 with the echelle grating and a 0\farcs9 slit. 
The two-pixel resolution, matching the slit width, was 
0.5 \AA\ (14 \kms). The stars were ratioed with comparison 
stars to remove telluric features. The flux calibrations were done using 
standards with colors from Koornneef (1984).
Wavelength calibration was achieved using
telluric absorption lines observed in the comparison stars. 
A quadratic fit was made to a selection of these lines covering the entire 
spectral range. The 1-$\sigma$ wavelength uncertainty is 
0.000005 -- 0.000008 \micron.

The observed sample is listed in Table 1. The spectra, slightly smoothed
to a resolution of 0.65 \AA\ (18 \kms), 
are shown in Figure 1.
The RCB stars range in effective temperature from 5000 to 7000 K.
Their spectra have very different appearances depending on whether the
RCB star is
warm (T$_{eff}$ = 6000--8000 K) or cool (T$_{eff}$ $<$ 6000 K)
(Asplund et al. 2000). 
In Figure 1,
the warm star spectra are in the lefthand column and the cool star spectra 
in the righthand 
column.
The warmer 
stars show mainly atomic absorptions of C I and singly ionized metals 
while the cooler 
stars, in addition, show strong C$_2$ and CN absorption bands. 
See Figure 2. The line 
identifications
are from Hirai (1974), and Hinkle, Wallace, \& Livingston (1995).

\section{Mass Outflow}

He~I $\lambda$10830 is present in the spectra of all of our
sample stars. It varies significantly in strength and shape, from a 
fully developed 
P-Cygni profile (ES Aql) to blue-shifted asymmetric absorptions with
small emission components, to even more complex structures.  
In all cases, the line indicates a mass outflow - with a range
of intensity and velocity. 
Given the temperatures of the RCB stars, it might be
expected that the photospheric
component of He~I $\lambda$10830 would be small.
However, absorption features of He~I $\lambda$5876 are present in several
RCB stars at velocities consistent with a photospheric origin (Rao \& 
Lambert 1996; Asplund et al. 2000).
A photospheric component of the He~I $\lambda$10830 line is not clearly
present in any star in our sample.

The P-Cygni-type line profiles in
our sample were modeled to obtain quantitative
information on the mass loss and the outflow velocity.
A detailed discussion of the individual objects is given below.
The results are compiled in Table \ref{tab_seifit} and shown in Figures
\ref{fig_seifit_ES} and  \ref{fig_seifit}.
Model profiles were computed with the SEI (``Sobolev plus Exact Integration'')
code, orginally developed by Lamers, Cerruti-Sola, \& Perinotto (1987) 
for wind lines
of hot stars in the UV range, and subsequently extended
to the analysis of H$\alpha$ lines (Bianchi et al. 1994). 
The code calculates the source function with the escape probability
method and the exact solution of the transfer equation. 
 We follow the notation used by Bianchi, Vassiliadas, \& Dopita (1997) and
Bianchi et al. (2000).
The velocity of the outflow increases outwards - following a law
characterized by the exponent $\gamma$ - 
until a terminal
velocity (\Vinf) is reached (eq. (1) of Bianchi et al. 2000).
In all profiles analyzed, we found a rather steep acceleration,
with  $\gamma$ $\approx$ 2. Terminal velocities are a few
hundred \kms.  
In many cases, the profiles are complex and blend with other photospheric 
absorption lines particularly in the cooler RCB stars,
making an accurate estimation of the parameters difficult.
In Figure 3, we show  an example of a profile that can be fit very
successfully using the SEI method. We also show, in Figure 4, other
profiles for which SEI is successful.
In some cases,
the He I absorption trough appears to be almost saturated, making the method
insensitive to the measurement of the total optical depth.
Several stars could not be fit successfully bacuse of saturated or complex
absorption profiles. 
The optical depth values obtained from the SEI analysis indicate
column densities of the He I outflow higher than 10$^{13}$ cm$^{-2}$ for 
all objects.

 ES Aql is the only member of the observed sample presenting
 a classical P-Cygni profile.  Its He I $\lambda$10830 spectral region is affected by
photospheric absorptions but much less than any other star in the sample.
Therefore, the location of the continuum and the fit of the He I line
profile were rather accurate. The best fit profile is
shown in Figure \ref{fig_seifit_ES} and the corresponding parameters 
 are given in Table \ref{tab_seifit}. The uncertainties in the total
optical depth and terminal velocity (\Vinf) were estimated by computing
several profiles, varying the parameters' around the best
fit solution, and narrowing the range of 
acceptable solutions considering both the 
S/N of the observed profile and the uncertainty in the continuum location.
 
In RY Sgr and R CrB, the He I absorption is much broader than the
emission, unlike in a classical P-Cygni profile.
The He I line profile fit is affected by two photospheric lines, 
Si I $\lambda$10830.1 and another line which may be a blend of 
S I $\lambda$10824.2 and Cr I $\lambda$10824.6 (See Figure \ref{fig_seifit}). 
The emission is partially masked by photospheric 
absorption.  
The derived parameters are much more uncertain in this case
than for the pure P-Cygni  profile of ES~Aql.
Two acceptable fits are shown for RY Sgr in Figure \ref{fig_seifit} - with
widely varying exponents of the optical depth law.
Therefore, 
the optical depth remains
very uncertain.

  WX CrA and  V517~Oph have very similar He I line profiles. 
The flat bottoms of the absorptions indicate that the line is saturated
and/or blended with other lines, both factors preventing an
accurate measurement of its optical depth.
The blue-shifted
part of the absorptions and the extreme red wings of the emission are 
well fit by P-Cygni profiles, provided a turbulence of about
20 - 30  \% of the terminal velocity is included (somewhat higher than the
10\% value which is typical for radiation pressure winds of hot stars). 
For these cooler RCB stars, many additional absorptions due to CN are 
obviously present, partially masking the emission. 
We nonetheless attempted to fit the blue wings of the
absorptions to obtain an estimate of the wind velocities.
Although the line shapes are similar for these two stars, the
velocity structures are different. The model profile adopted for
WX~CrA (\Vinf $\sim$225 \kms) also fits the blue wing of the 
V517~Oph (\Vinf $\sim$300 \kms) profile.  
However, the analysis indicates optical depths of  $\tau$ $>$1,
 and  velocities for the He I shells in the range found for other sample
objects. 

The line profiles of SV~Sge and  U~Aqr are also similar to each other, but 
in the spectrum of  U~Aqr the absorption is narrower,
implying a lower outflow velocity. The line has a smaller displacement 
blueward
than the previous cases.
U Aqr is located in the halo and has unusual abundances even for an RCB star
(Bond, Luck, \& Newman 1979).
The weakest He I features are seen in V854 Cen, V CrA and V482 Cyg. 
See Figure 1. 
The P-Cygni profile is present in V CrA with
the Si I line cutting through the middle of it. Also, V482 Cyg seems to
have its absorption hiding in a broadened Si I line. Only V854 Cen shows
little or no absorption, although there may be a weak broad absorption which
is very blue-shifted. 

\section{Discussion}

The lower state of the He I $\lambda$10830 transition is 20 eV above 
the ground state. This state is metastable as its transition probability is
very small (A$_{21}$=1.27 x 10$^{-4}$ s$^{-1}$) (Sasselov \& Lester 1994).  
It can be populated
by two mechanisms, photoionization/recombination or collisional excitation.
He II $\lambda$1640 is not seen in RCB stars 
(Clayton 1996; Lawson et al. 1999). 
This indicates that the He I $\lambda$10830 line is not being formed by
helium photoionization and recombination (Rossano et al. 1994).
Rao et al. (1999) observed the He I lines in emission 
at 3889, 5876, 7065 and 10830 \AA\
during the 1995-96 decline of R CrB.  
The 5876 \AA\ line was much weaker than the 3889 and 7065 \AA\ lines. 
This can be explained if the lines are optically thick and the electron 
density is high.
We cannot calculate density since we only have data on the 10830 line but
Rao et al. estimate T $\sim$ 20000 K and $n_e$ = 10$^{11}$--10$^{12}$ 
cm$^{-3}$.
In this regime, collisional excitation is important.
It has been suggested that the He I 
$\lambda$10830 line seen in 
Sakurai's object is
the result of collisional excitation in shocked gas being dragged outward by 
the expanding 
dust cloud
around the star (Eyres et al. 1999; Tyne et al. 2000).  
The dust formation and expansion by radiation pressure is 
thought to be 
quite similar in the RCB stars. 
In these fast-moving clouds, 
excitation might take place through atomic collisions or shocks (Feast 2001).

Blue-shifted high velocity absorption features (100-400 \kms) 
have been seen from time to time in the broad-line emission spectra of 
RCB stars both early in declines
and just before return to maximum light (Alexander et al. 1972; 
Cottrell, Lawson, \& Buchhorn 1990; 
Clayton et
al. 1992, 1993, 1994; Vanture \& Wallerstein 1995; Rao \& Lambert 1997; 
Goswami et al. 1997).
The blue-shifted absorptions can be understood if the
dust, once formed, is blown away from the star by radiation 
pressure, eventually dissipating and allowing the stellar photosphere to 
reappear.
The gas is dragged along with the dust moving away from the star.
The velocities seen in the P-Cygni profiles agree well with those measured
for the blue-shifted absorption features. 

There is a possible relationship between the lightcurve of an RCB
star, which represents the mass-loss history of the star, and the He I 
$\lambda$10830
profile. The last column of Table 1 lists the state of each RCB star when the 
He I spectra were obtained. None of the stars were in a deep decline but 
half of the stars (V517 Oph, WX CrA, ES Aql, SV Sge and
U Aqr)\footnote{These five stars also represent the cool RCB stars in our 
sample. 
By chance, these stars were all in decline while the warmer stars were not.
There is no known relationship between dust formation 
activity and effective temperature.
Future observations will determine if there is a dependence of the 
effective temperature of the star on the nature of the P-Cygni profiles.} 
were below maximum light and in the late stages of a decline. 
All of these stars show strong P-Cygni or asymmetric blue-shifted 
profiles. In addition, both R CrB and 
RY Sgr, which show fairly strong profiles, 
had just recently returned to maximum light. 
Two stars,
V CrA and V482 Cyg, show weaker profiles. These stars had been continuously 
at
maximum light for 800 and 1400 days, respectively, at the
time the spectra were taken.
V482 Cyg, which had gone the longest without a decline, has 
the weakest He I feature
in the sample.
 
An exception to this trend is V854 Cen.
Its spectrum shows no sign of He I
except possibly a weak broad absorption.
Like R CrB and RY Sgr, V854 Cen was
just out of a decline, so one might have predicted a strong profile.
V854 Cen is an unusual RCB star. It is extremely active.
Along with V CrA, which also has a weak He I line, V854 Cen is a member of
the minority abundance group of the RCB stars, and has a relatively high
hydrogen abundance (Asplund et al. 2000).
However, helium is
still the dominant element so it is not clear why the P-Cygni profile is so 
weak in
these stars.
V482 Cyg is not a member of this minority group.

Of the ten stars, only R CrB has been measured in the 
He I $\lambda$10830 line at more than one epoch. In addition to 
the June 2001 observation reported here, it was also observed in March and 
May 1972 (6 mag below maximum light), January 1978
(1 mag below maximum light),
July 1978 (at maximum light for 100 days), and May 1996 (3.5 mag below 
maximum light) (Wing et al. 1972; Querci \& Querci 1978, Zirin 1982, 
Rao et al. 1999).
The He I $\lambda$10830 line is present at all four epochs. It is a P-Cygni
profile in January 1978, a blue-shifted absorption in July 1978, 
a strong emission
line in May 1996 and a P-Cygni profile in June 2001. 
The value of \Vinf seems to be
fairly constant at $\sim$200-240 \kms. The P-Cygni or asymmetric blue-shifted 
profile in R CrB, at least, 
seems to be present at all times, whether the star is at maximum light 
or in decline.

The He I $\lambda$10830
line is a key to understanding the evolution of RCB stars and the nature of
their
of mass-loss.
The observed line profiles can help distinguish between the Final Flash 
and Double Degenerate models for the evolution of RCB stars.  
In the Double Degenerate scenario, an RCB star is unlikely to be a 
binary; thus the detection of a
companion would favor the Final Flash scenario.
Rao et al. (1999) have made the suggestion that RCB stars are 
binaries and that the higher temperature lines are formed in 
an accretion disk wind around a white dwarf companion. 
They suggest that  the He I lines may arise 
from the inner regions of the accretion disk while 
lower excitation lines such as the Na I D lines would
then arise in the outer parts of the disk.
No evidence for binarity has ever been found in the RCB stars. 
If the RCB stars
are single stars with mass-loss similar to Sakurai's object, 
we expect to see P-Cygni-type profiles, while if 
we 
are viewing an accretion
disk directly, we expect to see pure emission lines. 
A third possibility exists where 
the 
accretion disk is seen through the
material being lost by a cool companion. In this case, the profile would vary 
with 
orbital phase. 

In these He I profiles, we can see the mass-loss from the RCB 
stars for the first time. 
It has long been suggested that when dust forms around an RCB star, radiation 
pressure accelerates the dust away from the star dragging the gas along with 
it. 
But until now, we have only been able to measure the dust.
The He I $\lambda$10830 profiles measured here
allow us to study the velocity 
structure and optical depth of the gas escaping from the star.
We plan to monitor these stars to see how the 
column densities and velocities vary with time and how they are related to the
dust formation episodes. 

\acknowledgements
The United Kingdom Infrared Telescope is operated by the
Joint Astronomy Centre on behalf of the U.K. Particle Physics and Astronomy
Research Council. TRG's research is supported by the Gemini Observatory,
which is operated by the Association of Universities for Research in
Astronomy, Inc., on behalf of the international Gemini partnership. GCC 
appreciates
the hospitality of the Australian Defence Force Academy and the Mount Stromlo
and Siding Springs Observatories.
LB acknowledges support from NASA grant NRA-99-01-LTSA-029. We thank
Albert Jones for providing photometry of the RCB stars. We also thank the
referee for useful suggestions.

\begin{deluxetable}{lllllll}
\tablewidth{0pc} 
\tablenum{1}
\rotate
\tablecaption{Observed Stars}
\tablehead{ 
\colhead{Star} & \colhead{$V_{max}$ [mag]} & \colhead{T$_{eff}$\tablenotemark{a}} &
\colhead{$V_{\infty}$ [\kms]} &
\colhead{$\tau$} & \colhead{N$_{He I}$ [$\times$ 10$^{12}$ cm$^{-2}$]} & \colhead{Lightcurve Status\tablenotemark{b}}
}
\startdata 
ES Aql&11.7&Cool&355$\pm$15 & 1.8$\pm$0.2    &0.4$\pm$0.05&3 mag below max\\
RY Sgr&6.2&Warm&275$\pm$50 & 5-20 :         &2$\pm$1.5&just at end of last decline\\
U Aqr&11.2&Cool&175$\pm$40 & 2-5 :          &0.5$\pm$0.3&2 mag below max\\
V517 Oph&11.5&Cool&300:       & \nodata   & \nodata&0.5 mag below max\\
WX CrA&11.5&Cool&225$\pm$30 & $>$50\tablenotemark{c}       &$>>$1:&0.5 mag below max\\
SV Sge&11.0&Cool&230$\pm$30 & 5$\pm$3        &0.7$\pm$0.5&almost at end of last decline\\
R CrB&5.8&Warm&200: &\nodata&\nodata&100 d since end of last decline\\
VCrA&10.0&Warm&295:&\nodata&\nodata&800 d since end of last decline\\
V482 Cyg&11.1&Warm&260:&\nodata&\nodata&1400 d since end of last decline\\
V854 Cen&7.0&Warm&\nodata&\nodata&\nodata&just at end of last decline
\enddata
\tablenotetext{a}{Warm is defined as T$_{eff}$ = 6000--8000 K; Cool is defined as T$_{eff}$ = $<$6000 K.}
\tablenotetext{b}{Based on data from the American Association of Variable Star Observers.}
\tablenotetext{c}{ A very high optical depth is required to fit this line, making 
the method insensitive.}\label{tab_seifit}
\end{deluxetable}


\begin{figure*}
\epsscale{0.8}
\plotone{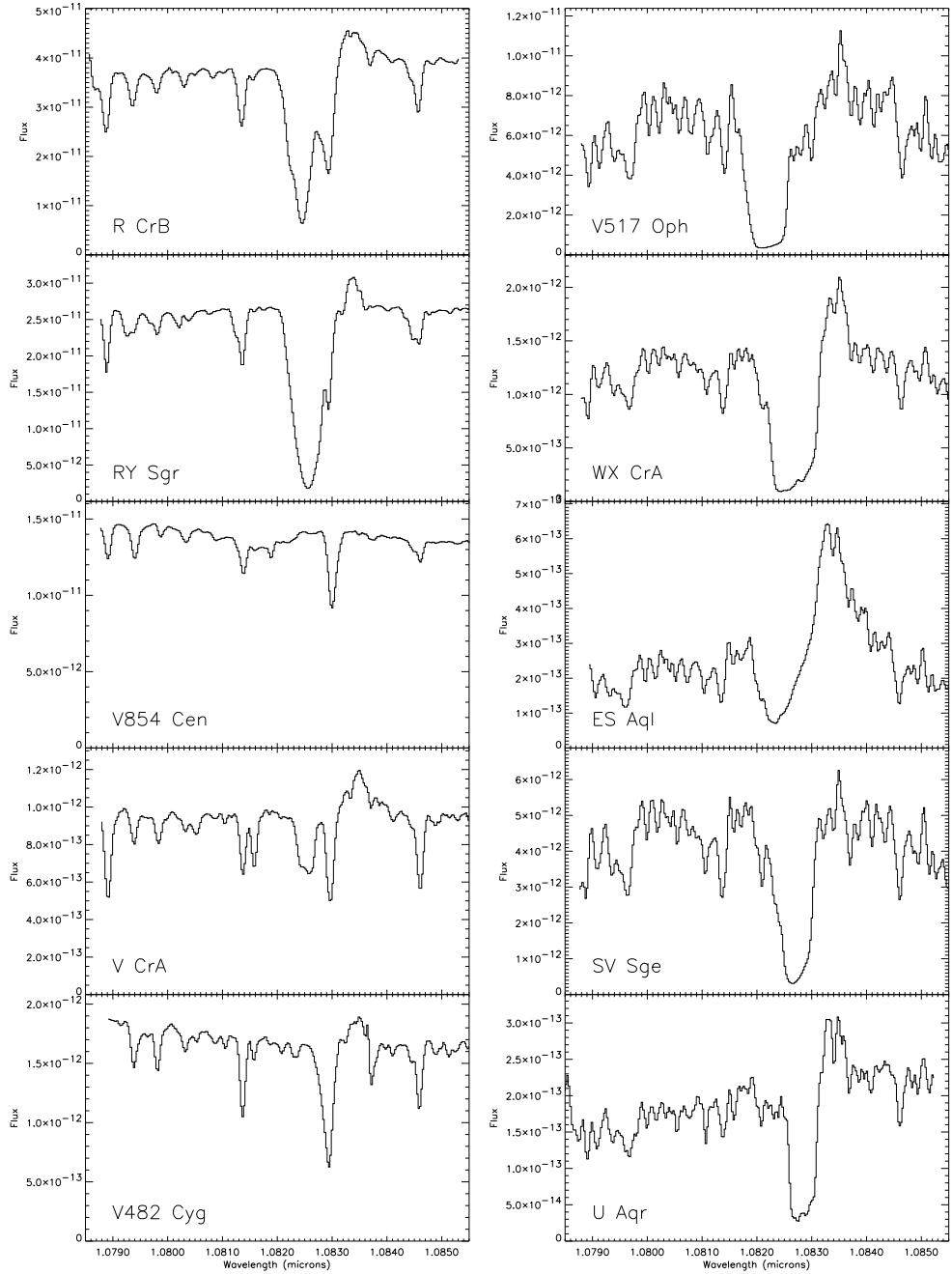}
\vspace{0.5 in}
\caption{UKIRT CGS4 echelle spectra of the He I $\lambda$10830 region for 
the sample 
RCB stars. The flux units are erg~cm$^{-2}$~s$^{-1}$~\AA$^{-1}$.
The plotted spectra have been smoothed with a Gaussian
of FWHM equal to 1.5 data points, so the resultant resolution is 18
km s$^{-1}$ (0.65 \AA).
Vacuum wavelengths are plotted.} \label{fig-2}
\end{figure*}

\begin{figure*}
\epsscale{0.8}
\plotone{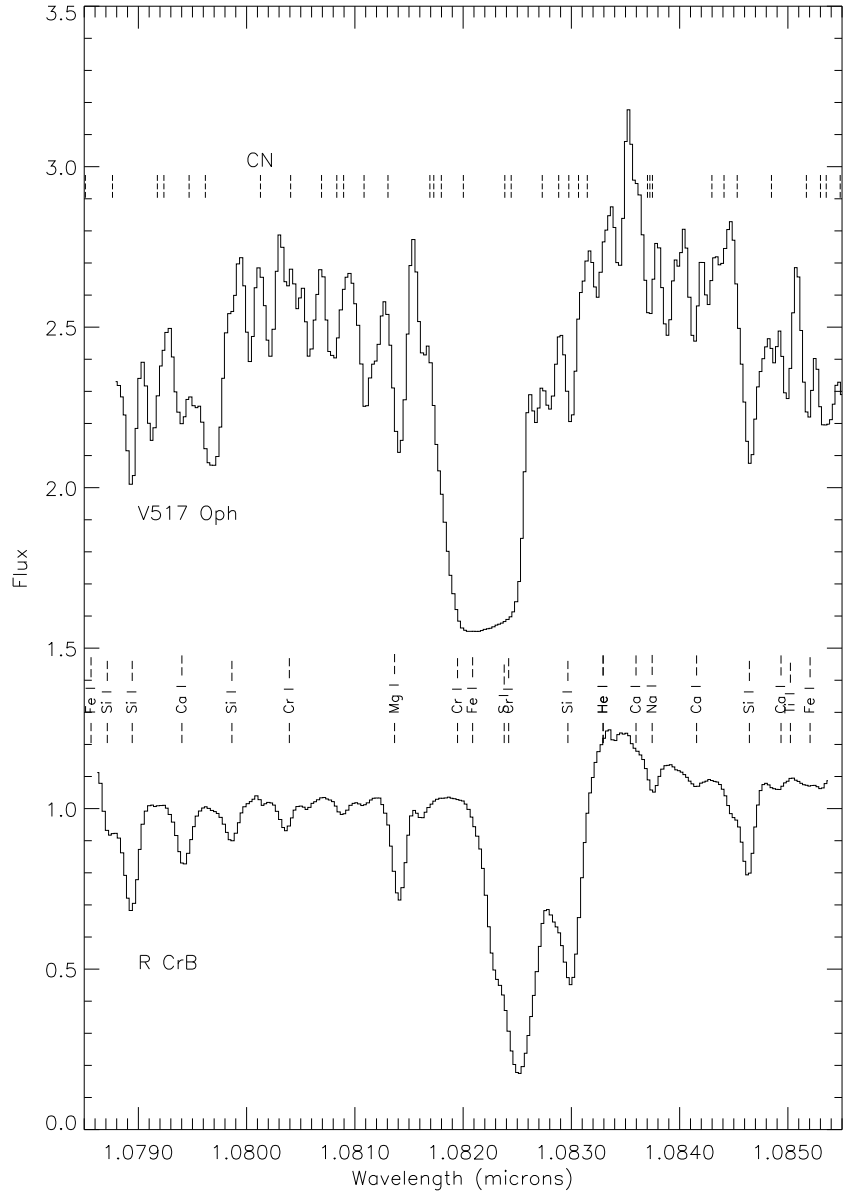}
\vspace{0.5 in}
\caption{Line identifications for two stars in the sample
showing typical examples of warm (R CrB) and cool (V517 Oph) RCB stars.
See text.
The flux units are arbitrary.
Vacuum wavelengths are plotted.} 
\label{fig-2}
\end{figure*}

\begin{figure*}
\epsscale{1.0}
\plotone{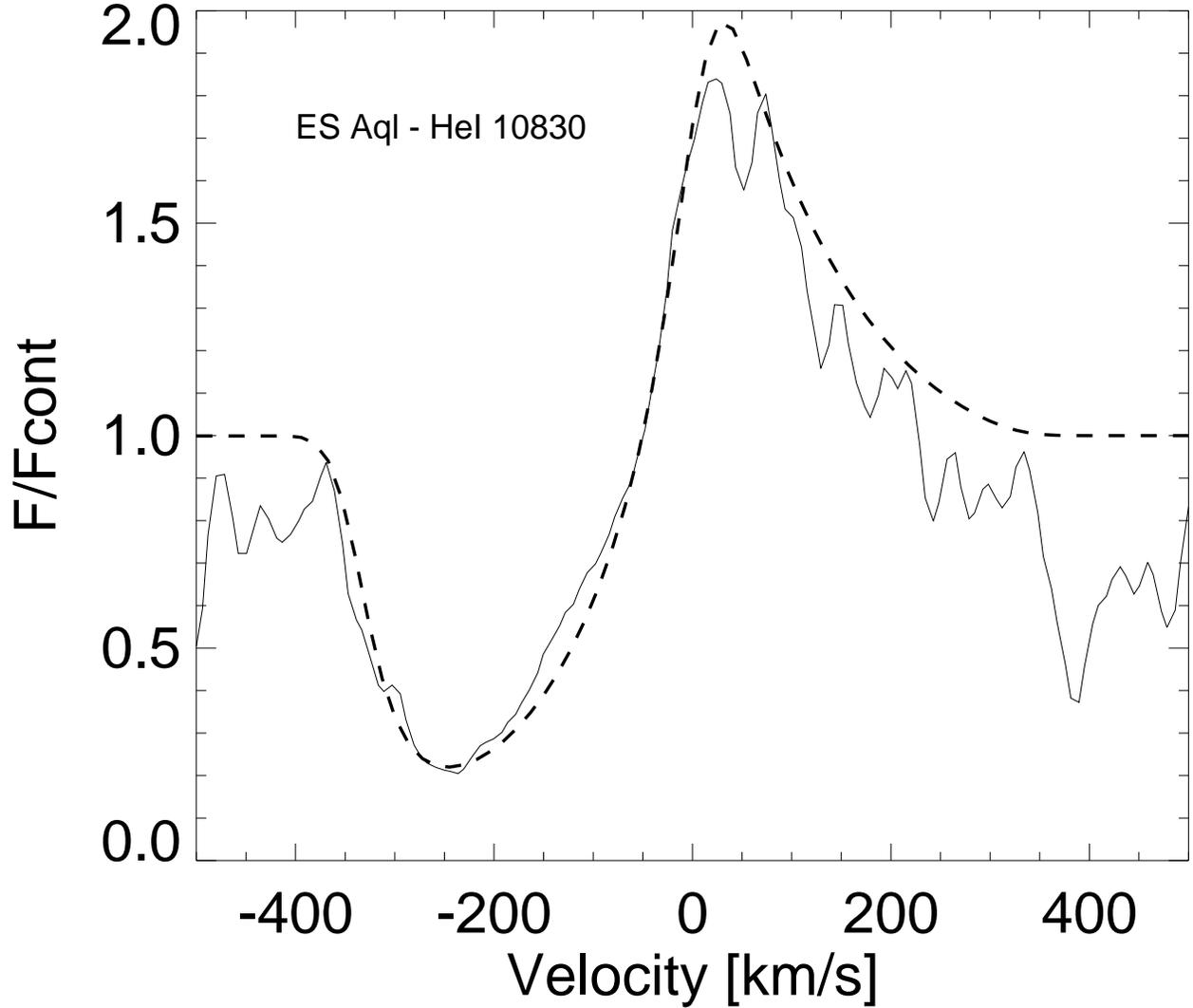}
\vspace{0.5 in}
\caption{The He~I 10830 line profile of ES~Aql is a classical
P-Cygni profile, revealing mass  outflow. The best fit
model profile (dashed line) provides a quantitative measurement
of the column density of He~I and of the outflow velocity
(see Table \ref{tab_seifit}). Zero velocity is the rest wavelength
of the transition, after the recession velocity of the star has been removed.
\label{fig_seifit_ES} }
\end{figure*}

\begin{figure*}
\epsscale{0.8}
\plotone{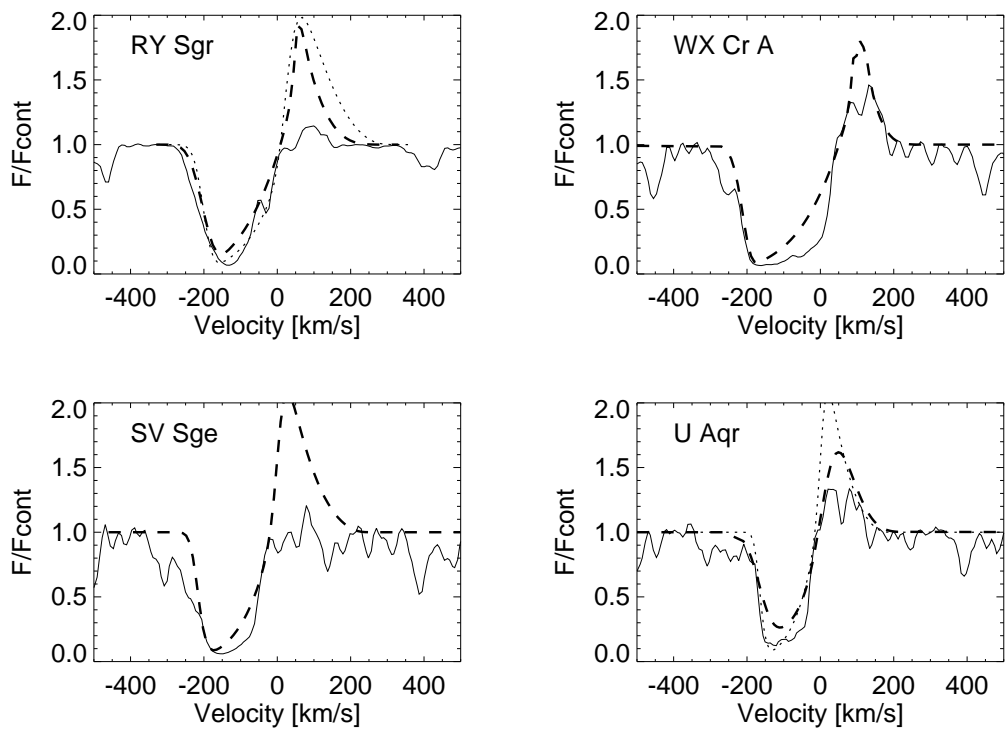}
\vspace{0.5 in}
\caption{The He I $\lambda$10830 line profiles of four stars in our sample
are shown, in velocity scale, with their best fit model profiles. 
The main absorption structure is asymmetric and blue-shifted, indicating
outflow, and the corresponding emission is probably masked by
other lines. The interpretation of which parts of the profiles
are the intrinsic He I line profiles becomes uncertain, and the two
possible fits shown for RY Sgr and U~Aqr show the range of 
uncertainty of the analysis in these cases. 
Zero velocity is the rest wavelength
of the transition after the recession velocity of the star has been removed.
\label{fig_seifit} }
\end{figure*}

\end{document}